# Interplay between Jahn-Teller instability, uniaxial magnetism and ferroelectricity in $Ca_3CoMnO_6$

Y. Zhang[1], H. J. Xiang[2] and M.-H. Whangbo[1]

[1] Department of Chemistry, North Carolina State University, Raleigh, NC 27695-8204

[2] National Renewable Energy Laboratory, 1617 Cole Blvd., Golden, CO 80401

**Abstract**

$Ca_3CoMnO_6$ is composed of $CoMnO_6$ chains made up of face-sharing $CoO_6$ trigonal prisms and $MnO_6$ octahedra. The structural, magnetic, and ferroelectric properties of this compound were investigated on the basis of density functional theory calculations. $Ca_3CoMnO_6$ is found to undergo a Jahn-Teller distortion associated with the $CoO_6$ trigonal prisms containing high-spin $Co^{2+}$ ($d^7$) ions, which removes the $C_3$ rotational symmetry and hence uniaxial magnetism. However, the Jahn-Teller distortion is not strong enough to fully quench the orbital moment of the high-spin $Co^{2+}$ ions thereby leading to an electronic state with substantial magnetic anisotropy. The Jahn-Teller distorted $Ca_3CoMnO_6$ in the magnetic ground state with up-up-down-down spin arrangement is predicted to have electric polarizations much greater than experimentally observed. Implications of the discrepancy between theory and experiment were discussed.

# I. Introduction

Orbital ordering in transition-metal magnetic oxides arises when their transition metal ions possess an electron configuration with unevenly filled degenerate d-states [e.g., three electrons in a doubly-degenerate d-state (xy, $x^2$-$y^2$) leading to the configuration (xy, $x^2$-$y^2$)$^3$] and hence Jahn-Teller (JT) instability [1]. Such an electron configuration is also required for a transition metal ion to have uniaxial magnetism [2], in which the ion has a nonzero magnetic moment only along the axis of the rotational symmetry (conventionally taken to be the z-axis) leading to the degenerate d-states, so that the spins of the ion become Ising spins. Thus, uniaxial magnetism is incompatible with JT instability, unless a JT distortion is prevented so that the degenerate d-states remain. For example, the high-spin $Fe^{2+}$ ($d^6$) ion at the linear two-coordinate site of $Fe[C(SiMe_3)_3]_2$ has the d-electron configuration (xy, $x^2$-$y^2$)$^3$(xz, yz)$^2$($z^2$)$^1$ [2, 3], so that $Fe[C(SiMe_3)_3]_2$ has JT instability but exhibits uniaxial magnetism because the sterically bulky $C(SiMe_3)_3$ groups prevent a JT distortion (i.e., the bending of the linear C-Fe-C framework). It is an open question what happens to uniaxial magnetism when a magnetic system with JT instability undergoes a weak JT distortion. This question is particularly relevant for the recently discovered multiferroic compound $Ca_3CoMnO_6$ [4], whose room-temperature crystal structure has the 3-fold rotational symmetry [5], because its high-spin $Co^{2+}$ ($d^7$) ions at the trigonal prism sites have the electron configuration ($z^2$)$^2$(xy, $x^2$-$y^2$)$^3$(xz, yz)$^2$ and hence have JT instability (see below). At present, it is unknown whether or not $Ca_3CoMnO_6$ undergoes a JT distortion at a low temperature, but its spins are regarded as Ising spins [4]. Thus, if a JT distortion takes place in $Ca_3CoMnO_6$, it will affect not only the magnetic anisotropy but also the ferroelectric (FE) polarization of $Ca_3CoMnO_6$.

In ferroelectrics driven by magnetism, the FE polarization arises from the loss of lattice inversion symmetry brought about by magnetic order [6, 7]. Experimentally, two types of magnetic order have been found to remove lattice inversion symmetry, namely, the spiral spin order in one-dimensional (1D) chains made up of identical magnetic ions [6 – 11] and the up-up-down-down (↑↑↓↓) spin order in 1D chains made up of two different magnetic ions alternating along the chain [4]. A number of magnetic oxides undergo spin spiral ordering and hence exhibit ferroelectricity [8 – 11]. So far, the FE polarization induced by ↑↑↓↓ spin order in 1D chains is found in only one example, i.e., $Ca_3CoMnO_6$ [4]. Recently, it has been proposed that helical spin order [12] and canted spin order [13] can also induce FE polarization.

$Ca_3CoMnO_6$ consists of $CoMnO_6$ chains in which face-sharing $CoO_6$ trigonal prisms and $MnO_6$ octahedra alternate along the c-direction with short Co-Mn distance (2.646 Å) (**Fig. 1a** and **1b**), and the Ca atoms surrounding the $CoMnO_6$ chains cap the edges of the $O_3$ triangles and the $O_4$ rectangles of the $CoO_6$ trigonal prisms (**Fig. 1c**) [5]. The magnetic susceptibility study of $Ca_3CoMnO_6$ suggested that the $MnO_6$ octahedra and the $CoO_6$ trigonal prisms have high-spin $Mn^{4+}$ ($d^3$) and high spin $Co^{2+}$ ($d^7$) ions, respectively [5]. In contrast, the neutron diffraction study reported that the $Co^{2+}$ ($d^7$) ions are in a low spin state [4]. On the basis of X-ray absorption spectroscopy and first principles density functional theory (DFT) calculations, Wu *et al.* showed unambiguously that the trigonal prism and octahedral sites of $Ca_3CoMnO_6$ are occupied by high-spin $Co^{2+}$ and high-spin $Mn^{4+}$ ions, respectively [14]. On lowering the temperature, $Ca_3CoMnO_6$ undergoes a long range magnetic ordering at 16.5 K, below which the spins of each CoMnO$_6$ chain adopts the ↑↑↓↓ arrangement thereby leading to FE polarization [4]. The spins of Ca$_3$CoMnO$_6$ are oriented parallel to the c-direction (//c) [4], and the high-spin Co$^{2+}$ ions possess a high orbital moment ($\mu_L$ = 1.7 $\mu_B$) according to Wu *et al.*'s DFT calculations [14]. These observations suggest that the Co$^{2+}$ ions have uniaxial magnetism. The high-spin Co$^{2+}$ (d$^7$) ion at a trigonal prism site of Ca$_3$CoMnO$_6$ has the configuration $(z^2)^2(xy, x^2-y^2)^3(xz, yz)^2$ [2] and hence JT instability. Unlike the case of Fe[C(SiMe$_3$)$_3$]$_2$, however, the CoO$_6$ trigonal prisms of Ca$_3$CoMnO$_6$ are not surrounded by bulky groups (**Fig. 1c**) so that their CoO$_6$ trigonal prisms may undergo a JT distortion hence removing the 3-fold rotational symmetry and lifting the degeneracy of the (xy, x$^2$-y$^2$) and (xz, yz) levels. The latter in turn will remove uniaxial magnetism, although a substantial magnetic anisotropy may remain in the JT distorted structure if the extent of JT distortion is small. Experimentally, the FE polarization of Ca$_3$CoMnO$_6$ is found only along the c-direction (i.e., P$_{//c}$ ≈ 90 $\mu$C/m$^2$ at 2 K) [4]. However, a JT distortion may give rise to a nonzero FE polarization perpendicular to the c-direction as well. In the following we probe these questions on the basis of DFT calculations.

## II. Calculations

First principles spin-polarized DFT calculations for Ca$_3$CoMnO$_6$ were performed using the projector augmented wave method encoded in the Vienna ab initio simulation package (VASP) [15] with the local density approximation (LDA). To properly describe the electron correlation associated with the 3d states of Co and Mn, we used the LDA plus on-site repulsion U (LDA+U) method of Dudarev *et al.* [16]. In addition, spin-orbit coupling (SOC) effects [17] were considered by performing LDA+U+SOC calculations with the spins oriented parallel to the c-direction (hereafter the //c-spin orientation) and also perpendicular to the c-direction (hereafter the ⊥c-spin orientation). Our LDA+U+SOC calculations were carried out with the energy criterion of $10^{-5}$ eV for self-consistency, the plane-wave cutoff energy of 400 eV, and a set of 3×3×3 k-points for the irreducible wedge of the Brillouin zone. Our LDA+U+SOC calculations with higher precision energy-convergence criterion (e.g., $10^{-8}$ eV for self-consistency) lead to the same results as obtained from the use of the energy-convergence criterion of $10^{-5}$ eV. In their study of $Ca_3CoMnO_6$ [14], Wu *et al*. carried out DFT+U+SOC calculations with the WIEN2k program package [15] using the generalized gradient approximation (GGA) [19] and the DFT+U method of Anisimov *et al* [20] (with U = 5.0 and 4.0 eV for Co and Mn, respectively, and J = 0.9 eV for both Co and Mn). In the DFT+U method of ref. 17 employed in our work, the on-site repulsion U represents the effective on site repulsion, which corresponds to U – J in the DFT+U method of ref. 20.

### III. Results and discussion

#### A. Magnetic ground state

To confirm that the ↑↑↓↓ state is the magnetic ground state, we consider four ordered spin states depicted in **Fig. 1d**, namely, the ↑↑↓↓ state, the antiferromagnetic state ↑↓↑↓, the ↓↑↑↑ state with identical spins at the Mn sites, and the ferromagnetic state ↑↑↑↑. For simplicity, the interchain interactions are regarded as ferromagnetic.

These states were constructed using the experimental geometry determined at room temperature [5]. Our LDA+U+SOC calculations for the experimental geometry of $Ca_3CoMnO_6$ using the VASP show that the ↑↑↓↓ and ↑↓↑↓ states are the two lowest-energy states for various values of U on Co and Mn. In their GGA+U+SOC calculations for $Ca_3CoMnO_6$ using the WIEN2k package [18], Wu *et al*. found that the state in which each $CoMnO_6$ chain has the ↑↓↑↓ spin arrangement is more stable than the state in which each $CoMnO_6$ chain has the ↑↑↓↓ spin arrangement. This result is not consistent with the experimental finding that the magnetic ground state of $Ca_3CoMnO_6$ has the ↑↑↓↓ spin arrangement in each $CoMnO_6$ chain [4]. In the initial stage of our DFT study, we also faced the same problem.

Our systematic LDA+U+SOC calculations for the experimental structure of $Ca_3CoMnO_6$ using the VASP reveal that the ↑↑↓↓ state becomes more stable than the ↑↓↑↓ state only when small U values are used for both Co and Mn ($U_{Co}$ and $U_{Mn}$, respectively), for example, $U_{Co} = U_{Mn} = 1.1$ eV, as summarized in **Table 1**. Thus, in our further LDA+U+SOC calculations using the VASP, we chose $U_{Co} = U_{Mn} = 1.1$ eV unless mentioned otherwise. Though small, these U values lead to an insulating gap for all ordered spin states of $Ca_3CoMnO_6$ considered in our work (see below). Note that, in reproducing the magnetic properties of $TbMnO_3$, it was found necessary to use a small U for Mn (i.e., 2.0 eV) [10]. It should be pointed out that our LDA+U+SOC calculations using the WIEN2k program package, in which all electrons are taken into consideration, a slightly larger U value (e.g., $U_{Co} = U_{Mn} = 2.1$ eV) can reproduce the experimental finding that the ↑↑↓↓ state is more stable than the ↑↓↑↓ state.

**B. Competition between uniaxial magnetism and Jahn-Teller instability**

**Fig. 2a** shows the plots of the partial density of states (PDOS) of the Co 3d states in the ground state ↑↑↓↓ obtained from LDA+U+SOC calculations for the experimental structure. The up- and down-spin $z^2$ states are both occupied. For each of the xy and $x^2$-$y^2$ states, the up-spin state is occupied while only one of the down-spin states (split due to SOC) is occupied. For each of the xz and yz states, the up-spin state is occupied but the down-spin state is not. Consequently, the electron configuration of each Co atom is best described as

$$(z^2\uparrow)^1 (xy\uparrow, x^2\text{-}y^2\uparrow)^2 (xz\uparrow, yz\uparrow)^2 (z^2\downarrow)^1 (xy\downarrow, x^2\text{-}y^2\downarrow)^1,$$

which corresponds to a high-spin $Co^{2+}$ ($d^7$) ion as anticipated. This configuration has three electrons in the doubly degenerate level, i.e., $(xy, x^2\text{-}y^2)^3$, so that the $Co^{2+}$ ions will lead to not only uniaxial magnetism [2] but also JT instability.

To see if $Ca_3CoMnO_6$ undergoes a JT distortion, the structure of $Ca_3CoMnO_6$ in each of the four ordered spin states was optimized by LDA+U+SOC calculations. In this optimization, the cell parameters were fixed to the experimental values, but the atom positions were allowed to relax. Our calculations give rise to two different optimized structures, namely, the optimized structure with high orbital moment on $Co^{2+}$ (i.e., 1.50 $\mu_B$) and that with low orbital moment (i.e., 0.56 $\mu_B$) on $Co^{2+}$ (hereafter referred to as the high-$\mu_L$ and low-$\mu_L$ optimized structures, respectively). Our analysis of the optimized structures show that the high-$\mu_L$ optimized structure keeps the 3-fold rotational symmetry, whereas the low-$\mu_L$ optimized structure does not [21]. The relative energies, in meV per formula unit (FU), of the four ordered spin states determined from our LDA+U+SOC calculations with spins oriented along the c-direction are summarized in **Table 2a**, which shows that the ↑↑↓↓ state is the ground state for both the experimental and the optimized structures. The energy of each state increases in the order, the experimental structure > the high-$\mu_L$ optimized structure > the low-$\mu_L$ optimized structure. Thus, for each state, the distortion toward the low-$\mu_L$ optimized structure is a JT distortion because it removes the $C_3$ rotational symmetry. As summarized in **Table 2b**, this conclusion remains valid when larger values of $U_{Co}$ and $U_{Mn}$ are used for LDA+U+SOC calculations.

In the ↑↑↓↓ state the Co-Mn distance is 2.599 Å when the spins of the two sites are identical, and 2.693 Å otherwise. The atom displacements involved in the JT distortions in the ↑↑↓↓ and ↑↑↑↑ states, with respect to the experimental structure, are depicted in **Fig. 3**, where the largest atom movement is 0.064 Å. An important consequence of the JT distortion is that the orbital moment of the $Co^{2+}$ ion is reduced by a factor of approximately three (i.e., from 1.50 $\mu_B$ to 0.56 $\mu_B$) (**Table 3**). The orbital moment of the $Co^{2+}$ ion is nonzero, which indicates that the JT distortion is not strong enough to completely quench the orbital angular momentum of $Co^{2+}$. This can be seen from the PDOS plots of the Co xy and $x^2$-$y^2$ states calculated for the ↑↑↓↓ state of $Ca_3CoMnO_6$ using the optimized structure with Jahn-Teller distortion (i.e., without $C_3$ rotational symmetry) shown in **Fig. 2b**. Unlike the case of the experimental structure (with $C_3$-rotation symmetry) (**Fig. 2a**), the two PDOS peaks for the xy or the $x^2$-$y^2$ state are not identical for the optimized structure with Jahn-Teller distortion (i.e., without $C_3$ rotational symmetry) (**Fig. 2b**). However, the difference between the two PDOS peaks is not strong.

### C. Magnetic anisotropy

In the presence of a JT distortion, $Ca_3CoMnO_6$ in the ↑↑↓↓ state cannot have uniaxial magnetism due to the loss of the 3-fold rotational symmetry [2]. The observed magnetic anisotropy of $Ca_3CoMnO_6$ indicates that, in the ↑↑↓↓ state of the JT-distorted structure, the spins still prefer to orient along the c-direction. To confirm this implication, we also carried out LDA+U+SOC calculations for the ↑↑↓↓ state of the JT-distorted $Ca_3CoMnO_6$ with spins oriented perpendicular to the c-axis (⊥c). These calculations show that the ⊥c-spin orientation is less stable than the //c-spin orientation by 6.9 meV/FU, and leads to a smaller orbital moment for $Co^{2+}$ than does the //c-spin orientation (i.e., 0.16 vs. 0.56 $\mu_B$). Thus, the spins of $Ca_3CoMnO_6$ prefer to orient along the c-direction even in the JT distorted structure, which renders the observed anisotropic magnetic character to $Ca_3CoMnO_6$. Similarly, LDA+U+SOC calculations for the isostructural magnetic oxides $Ca_3Co_2O_6$ and $Ca_3CoRhO_6$, which have high-spin $Co^{3+}$ and $Co^{2+}$ ions at the trigonal prism sites, respectively, show that they undergo a JT distortion thereby losing uniaxial magnetism but their spins strongly prefer to orient along the c-axis [22].

### D. Spin exchange interactions

To see why the ground magnetic state of $Ca_3CoMnO_6$ is the ↑↑↓↓ state, we analyze the three intra-chain spin exchange interactions, i.e., the superexchange (SE) interaction $J_{Co-Mn}$ between the nearest-neighbor $Co^{2+}$ and $Mn^{4+}$ ions, the super-superexchange (SSE) interaction $J_{Co-Co}$ between the two adjacent $Co^{2+}$ ions, and the SSE interaction $J_{Mn-Mn}$ between the two adjacent $Mn^{4+}$ ions. For a pair of spin sites interacting via the spin exchange parameter J, the spin exchange energy is given by $-mnJ/4$, where $m$ and $n$ are the numbers of unpaired spins at the spin sites [23]. For the high-spin $Co^{2+}$ and $Mn^{4+}$ sites of $Ca_3CoMnO_6$, that $m = n = 3$. Therefore, in terms of the Heisenberg spin Hamiltonian made up of the spin exchange parameters $J_{Co-Mn}$, $J_{Co-Co}$ and $J_{Mn-Mn}$, the total spin exchange energies per FU of the four ordered spin states in **Fig. 1c** are written as

$E_{\uparrow\uparrow\downarrow\downarrow} = (9/4)(-J_{Mn-Mn} - J_{Co-Co})$,

$E_{\uparrow\downarrow\uparrow\downarrow} = (9/4)(-2J_{Co-Mn} + J_{Mn-Mn} + J_{Co-Co})$,

$E_{\downarrow\uparrow\uparrow\uparrow} = (9/4)(J_{Mn-Mn} - J_{Co-Co})$,

$E_{\uparrow\uparrow\uparrow\uparrow} = (9/4)(2J_{Co-Mn} + J_{Mn-Mn} + J_{Co-Co})$.

By mapping the energy differences between the four states given in terms of the spin exchange parameters onto the corresponding energy differences obtained from the LDA+U+SOC calculations (**Table 2a**), we obtain the values of $J_{Co-Mn}$, $J_{Co-Co}$ and $J_{Mn-Mn}$ listed in **Table 4**. These parameters are all antiferromagnetic, and hence are not in support of the assumption by Choi et al. [4] that the SE interaction $J_{Co-Mn}$ is ferromagnetic. The SE interaction $J_{Co-Mn}$ is strongly antiferromagnetic due to the short Co-Mn distance, which allows direct metal-metal overlap. The three exchange parameters satisfy the condition, $J_{Mn-Mn} + J_{Co-Co} > J_{Co-Mn}$, which makes the ↑↑↓↓ state more stable than the ↑↓↑↓ state.

The strengths of these interactions decrease in the order $J_{Mn-Mn} > J_{Co-Mn} > J_{Co-Co}$. It is of interest to probe why the interaction $J_{Mn-Mn}$ is strong but the interaction $J_{Co-Co}$ is weak. The $Mn^{4+}$ ($d^3$) ion at an octahedral site has the $(t_{2g})^3$ configuration, and each $t_{2g}$ orbital has the Mn 3d orbital combined out-of-phase with the O 2p orbitals and is contained in the $MnO_4$ equatorial plane (**Fig. 4a**) [24]. As depicted in **Fig. 4b** and **4c**, the magnetic orbitals of two adjacent $Mn^{4+}$ sites can overlap well through their O 2p orbitals thereby leading to a strong antiferromagnetic interaction [24]. The SSE interaction $J_{Co-Co}$ is not strong because the magnetic orbitals of two adjacent $CoO_6$ trigonal prisms (**Fig. 4d**) cannot effectively overlap through the intervening $MnO_6$ octahedron because the rectangular faces of the two $CoO_6$ trigonal prisms faces are farthest away from each other (**Fig. 4e**).

It is noted from **Table 2** that the energy difference between the ↑↑↓↓ and other states becomes larger after geometry optimization, consistent with the exchange striction mechanism for the ferroelectricity in $Ca_3CoMnO_6$. In particular, in the ↑↑↓↓ state, the spin-up (spin-down) Co atom moves towards (away from) the spin-up Mn atom to break the inversion symmetry. These displacements of the Co atoms confirm those suggested by Choi *et al*. [4], who suggested that the bond between Mn and Co with opposite spins is elongated to minimize the exchange interaction energy. This suggestion is not consistent with the fact that $J_{Co-Mn}$ is antiferromagnetic rather than ferromagnetic. We attribute the displacements to the fact that direct metal-metal bonding is stronger between adjacent metal ions with identical spin than between adjacent metal ions with opposite spins [25].

**E. Ferroelectric polarization**

12To estimate the effect of the JT distortion on the FE polarization of $Ca_3CoMnO_6$ in the ↑↑↓↓ state, the electric polarizations for the optimized structures of $Ca_3CoMnO_6$ were calculated using the Berry phase method [26]. With this method, one calculates differences in polarization, so it is necessary to consider the "paraelectric" state to be used as the reference. It is difficult to identify an insulating paraelectric state for $Ca_3CoMnO_6$. However, we note that there exist two JT-distorted structures with the same total energy but with opposite polarizations with respect to the paraelectric state; one structure leads to the other when the Cartesian coordinate of each atom is inverted [i.e., (x, y, z) → (-x, -y, -z)]. Then, the FE polarization of the JT-distorted ground state is taken as one half of the difference between the polarizations of the two JT-distorted structures.

For the high-$\mu_L$ optimized structure of $Ca_3CoMnO_6$ (i.e., the structure with $C_3$ rotational symmetry), the polarization along the c-direction is nonzero ($P_{//c}$ = 17700 $\mu C/m^2$), but that perpendicular to the c-direction is zero ($P_{\perp c}$ = 0). For the low-$\mu_L$ optimized structure of $Ca_3CoMnO_6$ (i.e., the structure without $C_3$ rotational symmetry), $P_{\perp c}$ is nonzero and is greater than $P_{//}$ (i.e., 16800 vs. 12500 $\mu C/m^2$). It should be pointed out that our LDA+U+SOC calculations implicitly assumed an FE arrangement of the $\perp c$ electric polarizations of the $CoMnO_6$ chains; to reduce the computational task, our calculations for the ↑↑↓↓ state of $Ca_3CoMnO_6$ assumed the ferromagnetic arrangement between the CoMnO chains, which leads to the FE arrangement between the JT-distorted CoMnO chains.

Compared with the values of FE polarization calculated for the multiferroics such as $LiCuVO_4$, $LiCu_2O_2$ and $TbMnO_3$ [9 – 11] driven by spiral spin order, the calculated polarization for $Ca_3CoMnO_6$ is much larger than the experimental value ($P_{//c} \approx 90$ μC/m$^2$ and $P_{\perp c} = 0$) [4], and is only an order of magnitude smaller than the large polarization found for $LuFe_2O_4$ [27], which is driven by charge order. For $LiCuVO_4$ [9], $LiCu_2O_2$ [9] and $TbMnO_3$ [10, 11] as wells $LuFe_2O_4$ [27], the values of the polarizations obtained from DFT+U+SOC calculations are in reasonable agreement with experiment. To account for the unusually large difference between theory and experiment found for the FE polarization of $Ca_3CoMnO_6$, we note that $Ca_3CoMnO_6$ is a quasi one-dimension system in which the coupling between adjacent $CoMnO_6$ is weak. Therefore, it is probable that not all chains have the same direction of FE polarization even in a single-crystal sample thereby forming many different domains of FE polarization. In such a case, the measurements of FE polarization will lead to a small nonzero $P_{//c}$ and a vanishing $P_{\perp c}$ due to the cancellation of the FE polarizations from different domains. In our calculations, however, all chains are assumed to be in a FE arrangement (i.e., in a single FE domain).

## IV. Concluding remarks

In summary, the phenomena of JT instability, uniaxial magnetism and FE polarization are intimately related to each other in $Ca_3CoMnO_6$. At low temperature, $Ca_3CoMnO_6$ is predicted to undergo a JT distortion thereby losing uniaxial magnetism but retain substantial magnetic anisotropy with orbital moment 0.56 μ$_B$ along the chain direction. It would be of interest to verify the occurrence of a JT-distortion in Ca$_3$CoMnO$_6$, which should be detectable by Raman spectroscopy. For the JT distorted Ca$_3$CoMnO$_6$ in the magnetic ground state with ↑↑↓↓ spin order, our calculations predict a strong electric polarization, much larger than experimentally observed. This discrepancy can be explained if a single-crystal sample consists of many different domains with opposite FE polarization.

**Acknowledgments**

The work at North Carolina State University was supported by the Office of Basic Energy Sciences, Division of Materials Sciences, U. S. Department of Energy, under Grant DE-FG02-86ER45259.

**References**

[1] K. I. Kugel and D. I. Khomskii, Sov. Phys. Usp. **25**, 231 (1982); I. B. Bersuker, The Jahn-Teller Effect, Cambridge University Press, 2006.

[2] D. Dai and M.-H. Whangbo, Inorg. Chem. **44**, 4407 (2005).

[3] W. M. Reiff, A. M. LaPointe and E. H. Witten, J. Am. Chem. Soc. **126**, 10206 (2004).

[4] Y. J. Choi, H. T. Yi, S. Lee, Q. Huang, V. Kiryukhin and S.-W. Cheong, Phys. Rev. Lett. **100,** 047601 (2008).

[5] V. G. Zubkov, G. V. Bazuev,1 A. P. Tyutyunnik, and I. F. Berger, J. Solid State Chem. **160**, 293 (2001).

[6] S. W. Cheong and M. Mostovoy, Nat. Mater. **6**, 13 (2007).

[7] D. I. Khomskii, J. Magn. Magn. Mater. **306**, 1 (2006).

[8] T. Kimura, T. Goto, H. Shintani, K. Ishizaka, T. Arima and Y. Tokura, Nature

(London) 426, 55 (2003).

[9] H. J. Xiang and M.-H. Whangbo, Phys. Rev. Lett. 99, 257203 (2007).

[10] H. J. Xiang, Su-Huai Wei, M.-H. Whangbo, and Juarez L. F. Da Silva, Phys. Rev. Lett. 101, 037209 (2008).

[11] A. Malashevich and D. Vanderbilt, Phys. Rev. Lett. 101, 037210 (2008).

[12] T.-H. Arima, J. Phys. Soc. Jpn. **76**, 073702 (2007).

[13] T. A. Kaplan and S. D. Mahanti, arXiv: 0808.0336v3.

[14] H. Wu, T. Burnus, Z. Hu, C. Martin, A. Maignan, J. C. Cezar, A. Tanaka, N. B. Brookes, D. I. Khomskii and L. H. Tjeng, arXiv: 0806.1607v1.

[15] Kresse, G.; Hafner, J. *Phys. Rev. B* **1993**, *62*, 558; Kresse, G.; Furthmüller, J. *Comput. Mater. Sci.* **1996**, *6*, 15; Kresse, G.; Furthmüller, J. *Phys. Rev. B* **1996**, *54*, 11169.

[16] S. L. Dudarev, G. A. Botton, S. Y. Savrasov, C. J. Humphreys, A. P. Sutton, Phys. Rev. B **57**, 1505 (1998).

[17] K. Kuneš, P. Novák, M. Diviš and P. M. Oppeneer, Phys. Rev. B, **63**, 205111 (2001).

[18] P. Blaha et al., in WIEN2K, An Augmented Plane Wave Plus Local Orbitals Program for Calculating Crystal Properties, edited by K. Schwarz (Techn. Universität Wien, Austria, 2001).

[19] J. P. Perdew, K. Burke, and M. Ernzerhof, Phys. Rev. Lett. 77, 3865 (1996).

[20] V. I. Anisimov, I. V. Solovyev, M. A. Korotin, M. T. Czyżyk, and G. A. Sawatzky, Phys. Rev. B 48, 16929 (1993).

[21] The high orbital moment found for the $Co^{2+}$ ion (i.e., 1.7 $\mu_B$) by Wu et al. [11]

indicates that the structure of $Ca_3CoMnO_6$ employed for their WIEN2k calculations has the $C_3$ rotational symmetry.

[22] Y. Zhang, H. J. Xiang, E. J. Kan, A. Villesuzanne and M.-H. Whangbo, in preparation.

[23] D. Dai and M.-H. Whangbo, J. Chem. Phys. **114**, 2887 (2001); D. Dai and M.-H. Whangbo, J. Chem. Phys. **118**, 29 (2003).

[24] M.-H. Whangbo, H.-J. and D. Dai, J. Solid State Chem. **176**, 417 (2003).

[25] D. Dai, H. J. Xiang and M.-H. Whangbo, J. Comput. Chem. **29**, 2187 (2008).

[26] R. D. King-Smith and D. Vanderbilt, Phys. Rev. B **47**, 1651 (1993); R. Resta, Rev. Mod. Phys. **66**, 899 (1994).

[27] H. J. Xiang and M.-H. Whangbo, Phys. Rev. Lett. **98**, 246403 (2007).

Table 1. Energy difference ΔE between the ↑↓↑↓ and the ↑↑↓↓ states of $Ca_3CoMnO_6$, ΔE = $E_{↑↑↓↓}$ − $E_{↑↓↑↓}$, obtained from LDA+U+SOC calculations as a function of the on-site repulsions on the Mn and Co atoms ($U_{Mn}$ and $U_{Co}$, respectively).

| $U_{Mn}$ (in eV) | $U_{Co}$ (in eV) | ΔE (meV/FU) |
|---|---|---|
| 1.1 | 1.1 | -3.4 |
| 2.1 | 2.1 | 14.7 |
| 2.1 | 4.1 | 23.9 |
| 4.1 | 2.1 | 19.1 |
| 4.1 | 4.1 | 25.8 |
| 4.1 | 6.1 | 23.6 |
| 6.1 | 4.1 | 30.7 |

Table 2a. Relative energies (in meV/FU) of the four ordered spin states of $Ca_3CoMnO_6$ determined from LDA+U+SOC calculations $U_{Mn} = U_{Co} = 1.1$ eV

| Structure used | ↑↑↓↓ | ↑↓↑↓ | ↓↑↑↑ | ↑↑↑↑ |
|---|---|---|---|---|
| Experimental | 0.00 (59.2)[a] | 1.69 | 9.41 | 31.75 |
| Optimized with high $\mu_L$ | 0.00 (31.3)[a] | 14.38 | 21.87 | 53.12 |
| Optimized with low $\mu_L$ | 0.00 (0.00)[a] | 11.89 | 21.26 | 52.93 |

[a] The numbers in the parentheses refer to the relative total energy (in meV) of the ↑↑↓↓ state with respect to the energy of the optimized structure without $C_3$ symmetry.

Table 2b. Relative energies (in meV/FU) of the ↑↓↑↓ and ↑↑↓↓ states of $Ca_3CoMnO_6$ determined from LDA+U+SOC calculations with $U_{Mn} = U_{Co} = 2.1$ and 4.1 eV

| Structure used | $U_{Mn} = U_{Co} = 2.1$ eV | | $U_{Mn} = U_{Co} = 4.1$ eV | |
|---|---|---|---|---|
| | ↑↑↓↓ | ↑↓↑↓ | ↑↑↓↓ | ↑↓↑↓ |
| Experimental | 0.00 (48.0)[a] | -7.36 | 0.00 (38.7)[a] | -12.9 |
| Optimized with high $\mu_L$ | 0.00 (31.3)[a] | -0.99 | 0.00 (30.3)[a] | -10.4 |
| Optimized with low $\mu_L$ | 0.00 (0.00)[a] | -4.20 | 0.00 (0.00)[a] | -12.3 |

[a] The numbers in the parentheses refer to the relative total energy (in meV) of the ↑↑↓↓ state with respect to the energy of the optimized structure without $C_3$ symmetry.

Table 3. Spin and orbital moments of the $Co^{2+}$ and $Mn^{4+}$ ions of $Ca_3CoMnO_6$ in the ↑↑↓↓ state determined from LDA+U+SOC calculations with $U_{Mn} = U_{Co} = 1.1$, 2.1 and 4.1 eV

| Geometry used | Spin Moment | | Orbital Moment [a] | |
|---|---|---|---|---|
| | $Co^{2+}$ | $Mn^{4+}$ | $Co^{2+}$ | $Mn^{4+}$ |
| Experimental [b] | 2.48 | 2.61 | 1.50 | -0.02 |
| | 2.54 | 2.73 | 1.57 | -0.02 |
| | 2.64 | 2.93 | 1.66 | -0.02 |
| Optimized with high $\mu_L$ [b] | 2.49 | 2.59 | 1.50 | -0.02 |
| | 2.54 | 2.72 | 1.57 | -0.02 |
| | 2.64 | 2.93 | 1.66 | -0.02 |
| Optimized with low $\mu_L$ [b] | 2.49 | 2.59 | 0.56 | -0.02 |
| | 2.54 | 2.72 | 0.60 | -0.02 |
| | 2.64 | 2.93 | 0.64 | -0.02 |

[a] The positive and negative orbital moments mean that they are in the same and opposite directions to the spin moments, respectively.

[b] The results for $U_{Mn} = U_{Co} = 1.1$ eV are given in the first row, those for $U_{Mn} = U_{Co} = 2.1$ eV in the scond row, and those for $U_{Mn} = U_{Co} = 4.1$ eV in the third row.

Table 4. Spin exchange parameters (in meV) of $Ca_3CoMnO_6$ extracted from LDA+U+SOC calculations with $U_{Mn} = U_{Co} = 1.1$ eV

| Geometry used | $J_{Co-Mn}$ | $J_{Mn-Mn}$ | $J_{Co-Co}$ |
|---|---|---|---|
| Experimental | 3.34 | 2.09 | 1.63 |
| Optimized with high $\mu_L$ | 4.31 | 4.86 | 2.64 |
| Optimized with low $\mu_L$ | 4.56 | 4.72 | 2.47 |

**Figure captions**

Figure 1.    (a) Projection view of the crystal structure of $Ca_3CoMnO_6$ along the c-direction. (b) Perspective view of an isolated $CoMnO_6$ chain. (c) Arrangement of nine Ca atoms surrounding one $CoO_6$ trigonal prism, in which the edges of the $O_3$ triangles and the $O_4$ rectangles are capped by Ca. (d) Four ordered spin arrangements of a single $CoMnO_6$ chain, where the large and small circles represent the $Co^{2+}$ and $Mn^{4+}$ ions, respectively, and the up- and down-spins are represented by the absence and presence of shading, respectively.

Figure 2.    (a) PDOS plots of the Co 3d states calculated for the ↑↑↓↓ state of $Ca_3CoMnO_6$ using the experimental structure, which has the $C_3$ rotational symmetry. The PDOS plots for the xy and $x^2$-$y^2$ states are identical, and so are those of the xz and yz states. (b) PDOS plots of the Co 3d states calculated for the ↑↑↓↓ state of $Ca_3CoMnO_6$ using the optimized structure with Jahn-Teller distortion (i.e., without $C_3$ rotational symmetry). For simplicity, only the PDOS plots for the xy and $x^2$-$y^2$ states are shown. The energy on the horizontal axis is in units of eV, and the PDOS on the vertical axis in units of states/eV per formula unit (the positive and negative scale indicates the up- and down-spin states, respectively).

Figure 3.    Displacements of the atoms associated with the Jahn-Teller distortions in the (a) ↑↑↓↓ and (b) ↑↑↑↑ states of $Ca_3CoMnO_6$ with respect to their positions of the experimental structure.

Figure 4. (a) Magnetic orbital contained in one $MnO_4$ equatorial plane of an isolated $MnO_6$ octahedron. (b) $MnO_4$ equatorial planes of two adjacent $MnO_6$ octahedra sharing one face of the intervening $CoO_6$ trigonal prism. (c) Projection view of the two magnetic orbitals associated with the two $MnO_4$ equatorial planes in (b). (d) Magnetic orbital of an isolated $CoO_6$ trigonal prism. (e) Rectangular faces of two adjacent $CoO_6$ trigonal prisms sharing an equatorial plane of the intervening $MnO_6$ octahedron.

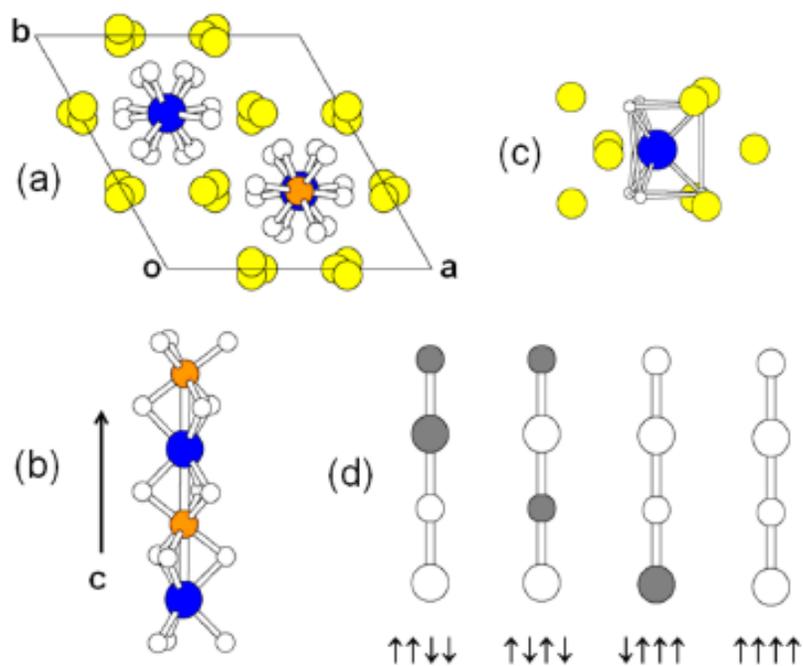

Figure 1.

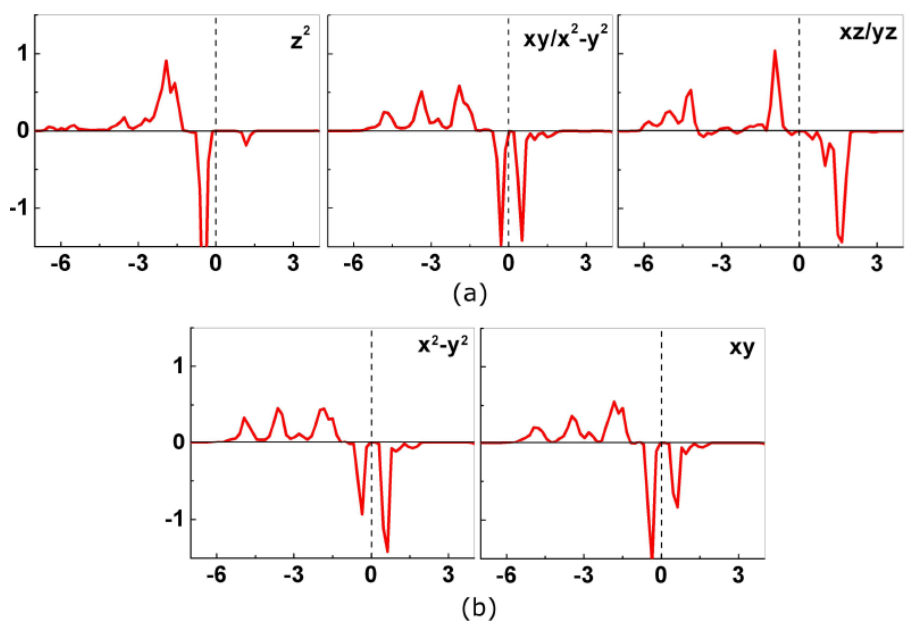

Figure 2.

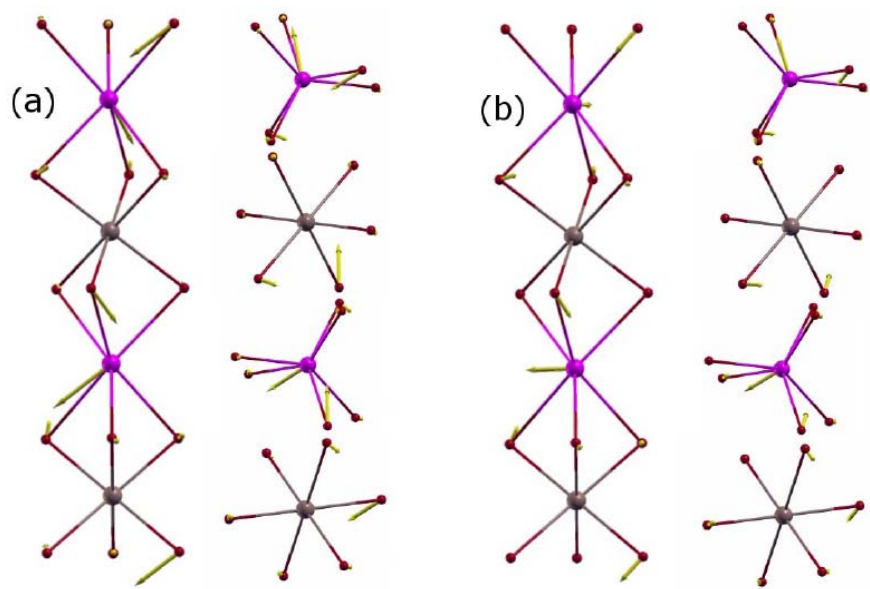

Figure 3.

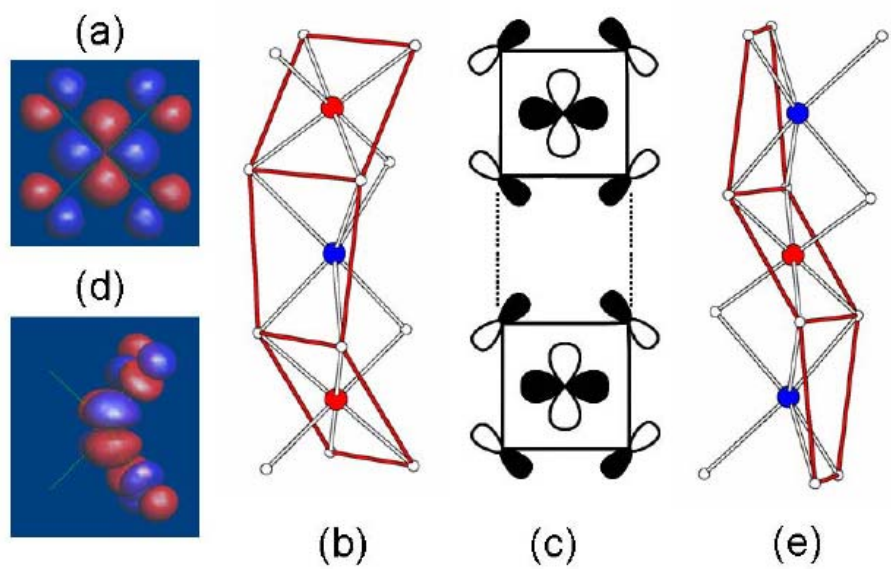

Figure 4.